\documentclass[12pt, usenatbib]{article}
\usepackage[margin = 1in]{geometry}
\usepackage{natbib}
\usepackage{bm}
\usepackage{amsmath, amssymb}
\usepackage{algorithm}
\usepackage{algpseudocode}
\usepackage{algorithmicx}
\usepackage{mathrsfs}
\usepackage{booktabs}
\newtheorem{example}{Example}
\newtheorem{lemma}{Lemma}
\usepackage{setspace}
\usepackage{pdfpages}

\newcommand{\Bernoulli}{\operatorname{Bernoulli}}

\newcommand{\Beta}{\operatorname{Beta}}
\newcommand{\Uniform}{\operatorname{Uniform}}
\newcommand{\Normal}{\mathcal N}
\newcommand{\Lognormal}{\log \Normal}
\newcommand{\Gam}{\operatorname{Gam}}

\newcommand{\Dirichlet}{\mathcal D}

\newcommand{\Halfcauchy}{\operatorname{half-Cauchy}}

\newcommand{\Var}{\operatorname{Var}}
\newcommand{\Cov}{\operatorname{Cov}}

\newcommand{\iid}{\stackrel{\textnormal{iid}}{\sim}}
\newcommand{\indep}{\stackrel{\textnormal{indep}}{\sim}}

\newcommand{\asim}{\stackrel{\text{\scalebox{2}{\(.\)}}}{\sim}}

\newcommand{\zeros}{\bm 0}

\newcommand{\bY}{\bm Y}
\newcommand{\bX}{\bm X}

\newcommand{\bx}{\bm x}

\newcommand{\sX}{\mathcal X}
\newcommand{\LPML}{\operatorname{LPML}}
\newcommand{\CPO}{\operatorname{CPO}}

\usepackage{hyperref}
\hypersetup{colorlinks=true,allcolors=blue}
\usepackage{hypcap}

\newcommand{\Tree}{\mathcal T}
\newcommand{\Leaves}{\mathcal L}

\newcommand{\sM}{\mathcal M}

\newcommand{\bomega}{\bm \omega}

\usepackage[figuresright]{rotating}

\title{Semiparametric Mixed-Scale Models Using Shared Bayesian Forests}

\author{
\normalsize  Antonio R. Linero \\
\normalsize  Department of Statistics \\
\normalsize  Florida State University \\
\normalsize  \texttt{arlinero@stat.fsu.edu}
\normalsize  \and 
\normalsize  Debajyoti Sinha \\
\normalsize  Department of Statistics \\
\normalsize  Florida State University \\
\normalsize  \texttt{sinhad@stat.fsu.edu}
\normalsize  \and 
\normalsize  Stuart R. Lipsitz \\
\normalsize  Department of Medicine \\
\normalsize  Brigham and Women's Hospital \\
\normalsize  \texttt{slipsitz@partners.org}
}

\begin{document}

% \date{{\it Received April} 2007. {\it Revised April} 2007.  {\it
% Accepted April} 2007.}

% \pagerange{\pageref{firstpage}--\pageref{lastpage}} 
% \volume{63}
% \pubyear{2007}
% \artmonth{December}

% \doi{10.1111/j.1541-0420.2005.00454.x}

% \label{firstpage}

%  put the summary for your paper here

\maketitle

\begin{abstract}
  This paper demonstrates the advantages of sharing information about unknown features of covariates across multiple model components in various nonparametric regression problems including multivariate, heteroscedastic, and semi-continuous responses. In this paper, we present methodology which allows for information to be shared nonparametrically across various model components using Bayesian sum-of-tree models. Our simulation results demonstrate that sharing of information across related model components is often very beneficial, particularly in sparse high-dimensional problems in which variable selection must be conducted. We illustrate our methodology by analyzing medical expenditure data from the Medical Expenditure Panel Survey (MEPS). To facilitate the Bayesian nonparametric regression analysis, we develop two novel models for analyzing the MEPS data using Bayesian additive regression trees - a heteroskedastic log-normal hurdle model with a ``shrink-towards-homoskedasticity'' prior, and a gamma hurdle model.

  \vspace{1em}
  
\end{abstract}

% \maketitle

\doublespacing

% \begin{keywords}
%   Bayesian additive regression trees;
%   Heteroskedastic errors;
%   Hurdle models;
%   Nonparametric Bayes;
%   Variable selection.
% \end{keywords}

% \maketitle

\section{Introduction}

\label{sec:introduction}

In complex statistical problems it is often of interest to share information across multiple model parameters and components. 
For studies with multiple responses, the same unknown set of features may be associated with the responses. 
In our motivating example of medical expenditure data from the Medical Expenditure Panel Survey (MEPS), many individuals record no medical expenditures (zero response) over the course of a year. As a consequence, the distribution of an individual's semi-continuous response of total yearly medical expenditures is a mixture of a point-mass at zero and a continuous distribution on the positive reals. Intuitively, the set of factors which predict whether an individual incurs \emph{no} medical expenditure may also be predictive of the \emph{magnitude}  that individual's medical expenditure if one occurs.

An increasingly popular method for modeling nonparametric functions is the Bayesian additive regression trees (BART) framework introduced by \citet{chipman2010bart}. The BART framework has been successfully applied to a diverse set of problems including survival analysis \citep{sparapani2016nonparametric}, causal inference \citep{hahn2017bayesian,hill2011bayesian}, analysis of loglinear models \citep{murray2017log}, imputation of missing predictors \citep{xu2016sequential}, and high dimensional prediction and variable selection \citep{linero2016bayesian}.

In this paper, we introduce \emph{shared forests}, which nonparametrically model multiple model components using a single set of trees. By viewing BART as a method for learning data-adaptive basis expansions, shared forests restrict the basis functions across model components to be the same while allowing the corresponding coefficients to be different. A simulation study shows that sharing information across model components in this fashion can be very beneficial, particularly in sparse high-dimensional problems in which variable selection a is necessary step.

In addition to our shared forests model, we make several additional contributions which are of practical interest in their own right. Semi-continuous responses are routinely modeled via two-part mixture models, often called hurdle models in econometrics, with a binary component modeling the probability of a zero response, and a continuous distribution modeling the response given it is non-zero. We present two novel semiparametric hurdle models for analyzing semi-continuous responses. The first is a type of gamma hurdle model, which are popular for modeling rainfall data \citep{feuerverger1979some}, in which the mean of the gamma distribution and the probability of a zero response are both modeled nonparametrically. 
The second model is a log-normal hurdle model \citep{aitchison1955distribution, xiao1999comparison} in which the log-mean, log-variance, and the probability of a zero response are all modeled nonparametrically. See \citet{tu2006zero} for a review of zero-inflated and hurdle models. To the best of our knowledge, we are first to adapt BART to the mean of a gamma distribution; this requires developing an analog of the usual Bayesian backfitting approach for fitting BART models of \citet{chipman2010bart}. Additionally, while nonparametric models for the variance have been considered in other Bayesian sum-of-trees approaches \citep{murray2017log, pratola2017heteroscedastic}, we are required to develop a different nonparametric approach for the log-variance of our log-normal hurdle model in order to allow the tree structures to be shared across the mean, variance, and probability of a zero response while preserving computational tractability. In order to prevent overfitting on the variance component of the model, our variance modeling framework is designed to be centered at, and to allow shrinking heavily towards, a parsimonious model with constant variance. This allows us to model heteroskedasticity in the data while preserving estimation efficiency when the variance of the response is actually constant.

Our approach has natural connections with several proposals in the machine learning literature on \emph{multi-task learning} and \emph{multi-output} learning; see \citet{borchani2015survey} for a review. Related methods include multi-objective decision trees, multi-task boosting, and multi-task kernel methods. 
Additionally, there are several methods which share information across models in the BART framework. 
The Bayesian causal forest (BCF) model of \citet{hahn2017bayesian} uses two separate forests to model the distribution of potential outcomes: the first forest accounts for the direct effect of confounders on the potential outcomes, and is identical for both the treatment and controls, while the second forest is unique to the distribution of the treated samples.
Another related work by \citet{starling2018functional} proposes a model in which a temporally indexed response is modeled using BART, with the forest being shared across time. We discuss these connections in Section~\ref{sec:related-methods}.

We apply our methodology to data from the Medical Expenditure Panel Survey (MEPS). 
The outcome $Y$ is an individual's total health care expenditure during the course of the year 2015. We show that the heteroskedastic log-normal hurdle model fits this data very well, and we use a shared forest to jointly model (i) the probability of $Y = 0$, (ii) the mean of $\log Y$ given 
$Y$ is nonzero, and (iii) the variance of $\log Y$ given $Y$ is nonzero. By examining the fit of the mean and variance components, we are able to validate the earlier observation of \citet{blough2000using} that the variance of $Y$ is roughly proportional to $E(Y)^{1.5}$ for MEPS data. However, we are also able to identify sources of heterogeneity which are not explained by this relationship between the variance and the mean.

In Section~\ref{sec:shared}, we introduce our shared forests framework. In Section~\ref{sec:models}, we develop and give default prior specifications for the gamma hurdle and log-normal hurdle models we use later to analyze the MEPS data. In Section~\ref{sec:simulation}, we conduct a simulation study which illustrates the potential benefits of sharing information across model components. In Section~\ref{sec:analysis}, we apply the methodology developed here to the MEPS dataset. We conclude in Section~\ref{sec:discussion} with a discussion. Additional computational details, and details about the analysis of the MEPS dataset, are given in a web appendix.

\section{Shared Forests}

\label{sec:shared}

\subsection{Review of Bayesian Additive Regression Trees}

\label{sec:review}

\begin{figure}
  \centering
  \includegraphics[width=1\textwidth]{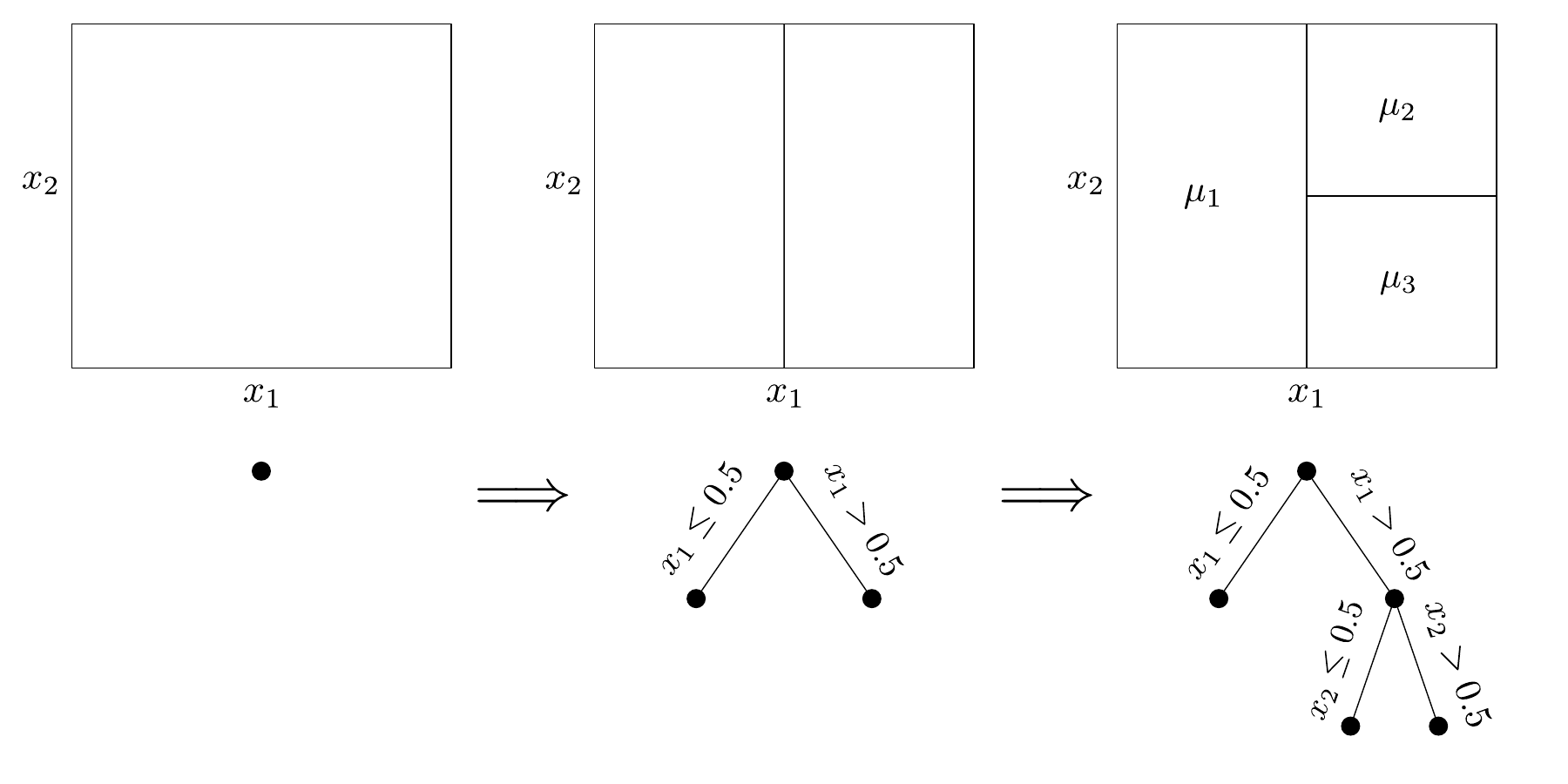}
  \caption{Schematic illustration of the construction of a decision tree (bottom) with the induced recursive partitioning of the predictor space $\sX = [0,1]^2$. After the decision tree is constructed, parameters associated to leaf node $\ell$ are given a mean parameter $\mu_{\ell}$.}
  \label{fig:GraphFigure}
\end{figure}

We briefly review the Bayesian Additive Regression Trees (BART) framework \citep{chipman2010bart}, an extremely useful tool for constructing highly flexible Bayesian semiparametric models. BART models typically outperform comparable linear models and often outperform machine learning techniques such as boosted decision trees and random forests.
We assume that the unknown function of interest $h(\bx)$ can be expressed as a sum of $T$ regression trees depending on tree structures $\Tree_t$ and leaf-node parameters $\sM_t$,
\begin{align}
\label{eq:basis}
  h(\bx) = \sum_{t=1}^T g(\bx; \Tree_t, \sM_t),
\end{align}
where $g(\bx; \Tree_t, \sM_t) = \mu_{t\ell}$ if the predictor value $\bx$ is associated to leaf node $\ell$ of tree $t$. As illustrated in Figure~\ref{fig:GraphFigure}, the decision tree $\Tree_t$ encodes a recursive partition of the predictor space $\sX = [0,1]^P$, with $g(\bx; \Tree_t, \sM_t)$ being piecewise-constant. Let $\Leaves_t$ denote the leaf nodes of the tree, so that $\sM_t = \{\mu_{t\ell} : \ell \in \Leaves_t\}$. 

The prior for $h(\cdot)$ in (\ref{eq:basis}) consists of a prior mass function $\pi_\Tree(\cdot)$ for the tree structures $\Tree_t$ and a prior on the leaf node parameters $\sM_t$. \citet{chipman2010bart} propose a branching process prior for $\Tree_t$.  A draw from this prior is obtained by generating, for each node at depth $d$, two child nodes with probability $q(d) = \gamma(1 + d)^{-\zeta}$; otherwise, the node becomes a leaf node (which defines a new equivalence class). This process iterates for $d = 0,1,2,\ldots$ until we reach a depth $d$ at which all of the nodes are leaves.
Note that $q(d)$ is not a mass function over $d$, but instead is the prior probability of a given leaf node being converted to a branch node. The case $\zeta = 0$ corresponds to the Galton-Watson process \citep{athreya2004branching}. A sufficient condition for termination of this branching process is $\zeta > 0$.
After the tree topology is generated, each branch node $b$ is associated to a decision rule of the form $[x_j \le C_b]$ where the coordinate $j\in \{1,\ldots,P\}$ is selected independently for each branch with probability $s_j$. 
Throughout, we will use the sparsity inducing Dirichlet prior $(s_1, \ldots, s_P) \sim \Dirichlet(\xi/P, \ldots, \xi/P)$ proposed by \citet{linero2016bayesian}. This prior concentrates on neighborhoods of sparse probability vectors, a fact which has been leveraged to perform variable selection in linear models \citep{bhattacharya2015dirichlet}, and adapt to irrelevant predictors in Gaussian process models \citep{bhattacharya2014anisotropic}. Intuitively, if $s_j$ is very small (e.g., $s_j < 10^{-10}$), then predictor $x_j$ is highly unlikely to appear within any splitting rule, effectively eliminating $x_j$ from the model. The Dirichlet prior encourages these extreme values of $s_j$, allowing the model to perform fully-Bayesian variable selection.
In this paper, we set $C_b \sim \Uniform(L_j,U_j)$ for the cut-points $C_b$ conditional on the tree topology, selected coordinate $j$, and the parameters of $b$'s ``ancestor nodes''.
Here $(L_1, U_1) \times \cdots \times (L_P, U_P)$ is the hyperrectangle defined by all values of $\bm x$ which lead to branch $b$. 

Let $\bomega$ be a vector of non-tree-specific parameters, such as the variance $\sigma^2 = \Var(Y_i \mid \bX_i)$ for a regression model with constant variance. Our model for the response $Y_i$ in this setting is expressed as $(Y_i \mid \bX_i = \bx, h, \bomega) \sim f\{y \mid h(\bx), \bomega\}$ where $\{f(\cdot \mid \mu, \bomega)\}$ is a parametric family. Conditional on the trees $\Tree_1,\ldots,\Tree_T$, the leaf node parameters $\{\mu_{t\ell} : \ell \in \Leaves_t, 1 \le t \le T\}$ are given iid priors $\mu_{t\ell} \sim \pi_\mu$. Usually $\pi_\mu$ is chosen to ensure that the integrated likelihood
\begin{align}
  \label{eq:integrated-like}
  \Lambda(\Tree_t)
  =
  \pi_\Tree(\Tree_t)
  \int
  \prod_{i = 1}^n f\{Y_i \mid h(\bX_i), \bomega\} \,
  \prod_{\ell \in \Leaves_t} \pi_\mu(\mu_{t\ell}) \ d\mu_{t\ell}
\end{align}
has a closed form expression. For example, in the regression setting, 
a popular choice for $\pi_{\mu}$ is the $ \Normal(0, \sigma^2_\mu)$ density. When using Markov Chain Monte Carlo (MCMC) to conduct Bayesian inference, $\Tree_t$ can be updated using Metropolis-Hastings, with $\Lambda(\Tree_t)$ used to compute the acceptance probability; see \citet{pratola2016efficient} for further details. In the original paper of \citet{chipman2010bart}, both $\Lambda(\Tree_t)$ and the full conditionals for the $\mu_{t\ell}$'s are calculated using Bayesian backfitting.

A large reason for the success of BART is the existence of highly effective ``default'' priors which can be expected to provide a reasonable baseline level of performance without requiring tuning by the user. As a default, we set $\gamma = 0.95$, $\zeta = 2$, and $\xi/(\xi + P) \sim \Beta(0.5,1)$; other prior specifications are model specific. Additionally, BART models have highly desirable theoretical properties \citep{linero2017abayesian, rockova2017posterior}; in particular, in regression problems, certain BART models attain near-minimax optimal rates of estimation for functions $h(\bx)$ with low-order interactions. For an in-depth review of Bayesian regression tree methods, see \citet{chipman2013bayesian} and \citet{linero2017review}.

A fact which will be useful for specifying priors later, and for making
connections with other approaches, is that, under the conditions $E(\mu_{t\ell})
= 0$ and $\Var(\mu_{t\ell}) = \sigma^2_{\mu} / T$, the prior converges to a
Gaussian process as $T \to \infty$. To see this heuristically, note that we can
write $h(\bx) = T^{-1/2} \sum_{t = 1}^T \mu^\star_{t\ell}$ where the
$\mu^{\star}_{t\ell}$'s are iid with mean $0$ and variance $\sigma^2_\mu$. In
general, for fixed $\bx, \bx'$, we have $\Cov\{h(\bx), h(\bx')\} = \sigma^2_\mu
\Pr(\bx \sim \bx')$ where the event $[\bx \sim \bx']$ denotes that $\bx$ and
$\bx'$ share the same terminal node in $\Tree_1$. An application of the
multivariate central limit theorem then establishes convergence of the finite
dimensional marginals to a multivariate normal
distribution. 
A natural alternative method is to simply apply Gaussian processes in practice.
BART has both practical and theoretical benefits over Gaussian process models.
First, the computational complexity of Gaussian process methods typically is
$O(n^3)$ where $n$ is the number of samples; conversely, in practice, BART has
computations which scale slightly faster than $O(nT)$ \citep{chipman2010bart}.
Second, the recent works of \citet{rockova2017posterior} and
\citet{linero2017abayesian} show that BART models are capable of adapting to
low-order interactions in the covariates. Third, as we will see in
Section~\ref{sec:simulation}, empirically, BART with a finite $T$ tends to
perform better than the limiting model as $T \to \infty$.

\subsection{The shared forests model}

\label{sec:multi}

We consider a generalization of (\ref{eq:basis}) and set
\begin{math}
  (Y_i \mid \bm X = \bm x, \bm h, \bomega)
  \sim f\{y \mid \bm h(\bx), \bomega\},
\end{math}
where $\bm h= (h_1, \ldots, h_M)$ is a collection of $m$ functions and each $h_m(\bx)$ 
is modeled non-parametrically as a sum of regression trees. Note that, as illustrated in Figure~\ref{fig:GraphFigure}, we are effectively modeling $h_m(\bx)$ in terms of a basis function expansion from an overcomplete family of basis functions
\begin{align*}
  h_m(\bx) = \sum_{t = 1}^T \sum_{\ell \in \Leaves_t^{(m)}} \mu^{(m)}_{t\ell} \psi^{(m)}_{t\ell}(\bx),
  \qquad
  \qquad
  \psi^{(m)}_{t\ell}(\bx) = I[\bx \leadsto (t,\ell)],
\end{align*}
where $[\bx \leadsto (t,\ell)]$ occurs if $\bx$ is associated to leaf $\ell$ of tree $\Tree_t^{(m)}$ and $I(A)$ is the indicator that the event $A$ occurs. We can then view $\{\psi_{t\ell}^{(m)} : 1 \le t \le T, 1 \le m \le M, \ell \in \Leaves_t^{(m)}\}$ as a collection of features which are adaptively learned from the data to approximate $\{h_1(\bx), \ldots, h_M(\bx)\}$.

Our proposed shared forest framework assumes that these basis functions are shared across $M$ model components; that is, we assume $\psi^{(m)}_{t\ell}(\bx) \equiv \psi_{t\ell}(\bx)$ for $m=1,\cdots,M$. Equivalently, we assume that the features which are useful for approximating $h_m(\bx)$ are the same features that are useful for approximating $h_{m'}(\bx)$.
Note, however, that a given feature $\psi_{t\ell}(\bx)$ has
different unknown coefficients (effects) $\mu_{t\ell}^{(1)}$ and  
$\mu_{t\ell}^{(2)}$ respectively for $h_1(\bx)$ and $h_2(\bx)$. 

This shared basis function framework is imposed by assuming that the $h_m(\bx)$'s are modeled using $T$ regression trees with the same collection of trees for all $M$ model components that are potentially affected by the covariate vector $\bx$. That is, we assume
\begin{align}
\label{eq:newtree}
  h_m(\bx) = \sum_{t=1}^T g(\bx ; \Tree_t, \sM_t^{(m)}) \ ,
\end{align}
where $\bm \mu_{t\ell} = (\mu_{t\ell}^{(m)} : 1 \le m \le M)$. We will assume the multivariate prior density $\bm \mu_{t\ell} \sim \pi_{\bm \mu}$, potentially allowing dependence across parameters $\mu_{t\ell}^{(m)}$ for different values of $m$. 

\begin{example}
  \label{ex:mixed}
  Consider a mixed-scale response $\bY_i = (Y_{i1}, Y_{i2}, Z_i)$ in which $Y_{ij} = h_j(\bx) + \epsilon_j$ where $\bm \epsilon \sim \Normal(0, \bm \Sigma)$, $Z_i \sim \Bernoulli[\Phi\{h_3(\bx)\}]$, and $\bm h$ is modeled with a shared forest with $\bm \mu_{t\ell} \sim \Normal(\zeros, \Sigma_\mu / T)$. We consider a variant of this model in Section~\ref{sec:simulation}. 
\end{example}

\begin{example}
  \label{ex:semi}
  Consider a semicontinuous response $(Y_i \mid \bX_i = \bx, \bm h, \bomega)$ where $Y_i > 0$ occurs with probability $\Phi\{h_1(\bx)\}$ and $(Y_i \mid Y_i > 0, \bm X_i = \bx, \bm h, \bomega) \sim \Gam[\alpha, \alpha \exp\{h_2(\bx)\}]$. We refer to this model as the gamma hurdle model; see Section~\ref{sec:gamma}. 
\end{example}

Bayesian inference for the shared forest model of (\ref{eq:newtree})  can be conducted by extending \eqref{eq:integrated-like} to incorporate priors on the parameters for the leaf nodes across the multiple model components giving the integrated likelihood 
\begin{align}
  \label{eq:integrated-multi}
  \begin{split}
  \Lambda(\Tree_t)
  &=
  \pi_\Tree(\Tree_t)
  \int
  \prod_{i = 1}^n f\{Y_i \mid \bm h(\bX_i), \bomega\} \,
  \left[ 
  \prod_{\ell \in \Leaves_t} \pi_{\bm \mu}(\bm \mu_{t\ell}) \ d\bm\mu_{t\ell}
  \right]
 \\&=
  \pi_\Tree(\Tree_t)
  \prod_{\ell \in \Leaves_t}
  \int
  \prod_{i: \bX_i \leadsto (t,\ell)}
  f\{Y_i \mid \bm h(\bX_i), \bomega\} \,
  \pi_{\bm \mu}(\bm \mu_{t\ell}) \ d\bm \mu_{t\ell}\ .
\end{split}
\end{align}
As before, if \eqref{eq:integrated-multi} has a  closed form then one can update $\Tree_t$ within an MCMC algorithm using standard Metropolis-Hastings proposals. 

\subsection{Related methods}

\label{sec:related-methods}

There are several proposals for BART models which are related to our shared forests model. \citet{hahn2017bayesian} consider a related structure in the context of causal inference; given a binary treatment $z$, they model potential outcomes $Y_i(z)$ as $Y_i(z) = h(\bX_i) + z\alpha(\bX_i) + \epsilon_i$, with both $h(\bx)$ and $\alpha(\bx)$ modeled using BART priors. This is referred to as a \emph{Bayesian causal forest} (BCF). The function $h(\bx)$ represents the prognostic effects of the covariates $\bX_i$, which is shared across both potential outcomes, while $\alpha(\bx)$ represents a treatment-covariate interaction, which is unique to the treated group. This differs from the shared forests framework we present in that we only require sharing the tree topologies across model components, while the BCF model shares the entire function $h(\bx)$. Alternatively, one may view the BCF model for $h(\bx)$ as a shared forest in which the values in leaf $\ell$, given by $(\mu_{\ell,0}, \mu_{\ell,1})$, are perfectly correlated. In the context of causal inference, this separation of the effect into a perfectly-shared forest $h(\bx)$ and a completely separate treatment effect $\alpha(\bx)$ is desirable because it gives the analyst a great deal of control over the prior information expressed about individual-level treatment effects. 

In the context of functional regression, \citet{starling2018functional} model a temporally-observed response using a BART model as $Y_i(t) = h_t(\bX_i) + \epsilon_{i}(t)$ where here $t \in \mathscr T$ indexes the observation time. The parameters of the leaf nodes of the trees in the ensemble are then modeled as random functions $\mu_\ell(t)$, with Gaussian process priors. The distinction between how $t$ and $\bx$ are incorporated into their model is referred to as \emph{targeted smoothing}, as the model induces a higher degree of smoothing over $t$ than $\bx$. This approach can also be cast as a type of shared forest model in which the collection of regression functions $\{h_t(\bx) : t \in \mathscr T\}$ share the same tree topology.  The dependence between $h_{t}(\bx)$ and $h_{t'}(\bx)$ induced using Gaussian processes is analogous to using the multivariate normal prior described in Example~\ref{ex:mixed}.

Shared forests have natural connections to many proposals for \emph{multi-task} or \emph{multi-output} methods in machine learning. The most immediate connections are with multi-objective decision trees (MODTs) initially proposed by \citet{de2002multivariate}. MODTs are grown in a CART-like fashion, but use a multivariate purity function for evaluating the quality of splits. In this way, splits are useful for predicting all outputs simultaneously. Our shared forests model with $T = 1$ is essentially a Bayesian version of a MODT, as the marginal likelihood \eqref{eq:integrated-multi} will be large when $\Tree_t$ gives good predictions across all tasks. MODTs can be ensembled in the usual ways via the bagging and random forests algorithms \citep{kocev2007ensembles}.

Our characterization of the shared forests model in terms of a shared basis function expansion is similar to the assumption of the multi-task feature learning approach of \citet{argyriou2007multi}. Our approach can also be related to the FIRE algorithm for fitting rule ensembles proposed by \citet{aho2012multi}. Each $\psi_{t\ell}(\bx)$ in the ensemble is a ``rule'' and the $\mu_{t\ell}$'s are task-specific weights assigned to each rule.  Additionally, in the same way that BART is analogous to gradient boosted decision trees \citep{chipman2010bart, freund1999short}, the shared forests model is analogous to the boosted multi-task learning approach of \citet{chapelle2011boosted}.

Using the connection between BART and Gaussian processes, we can also interpret the shared forests model in terms of multi-task kernel methods. Recall from  Section~\ref{sec:review} that BART can be thought of as approximately implementing Gaussian process regression when the number of trees $T$ is large, with kernel $K(\bx,\bx') = \sigma^2_\mu\Pr(\bx \sim \bx')$. If the leaf nodes of the ensemble across the tasks are endowed with the prior $\bm \mu_{t\ell} \sim \Normal(0, \Sigma_\mu / T)$ then the cross-task kernel function is given by $\Cov\{h_j(\bx), h_k(\bx)\} = \Sigma_{ij} \Pr(\bx \sim \bx')$. This matches the proposal of \citet{bonilla2008multi} for multi-task Gaussian processes. Sharing of information across tasks for the shared forest occurs at a deeper level still, however, as in the finite-$T$ case the rule-sharing interpretation of our approach still applies even if $\Sigma_{ij} = 0$; as we show in the simulation study of Section~\ref{sec:simulation}, substantial gains are possible even with $\Sigma_{ij} = 0$.

\section{Models for semicontinuous data}

\label{sec:models}

\subsection{Probit-based hurdle models}

Motivated by the MEPS dataset, we present two models for analyzing zero-inflated responses. Throughout, let $\pi(\bx) = \Pr(Y_i > 0 \mid \bX_i = \bx, \bm h, \bomega)$ denote the probability of a non-zero response. The gamma hurdle and log-normal hurdle models below are special cases of the probit-based hurdle model, where
\begin{math}
  \pi(\bx) = \Phi\{\theta_0 + h_\theta(\bx)\},
\end{math}
and 
\begin{math}
  (Y_i \mid Y_i > 0, \bX_i = \bx, \bm h, \bomega)
  \sim f\{y \mid \bm h_u(\bx), \bomega\}.
\end{math}
Here, $\{f(\cdot \mid \mu, \bomega)\}$ is a parametric family of densities for the positive part of $Y_i$. We model $\bm h = (h_\theta, \bm h_u)$ with a shared forest. Let $\theta_{t\ell}$ denote the parameter associated to leaf $\ell$ of $\Tree_t$ for $h_\theta$ and $u_{t\ell}$ the parameter associated to leaf $\ell$ of $\Tree_t$ for $\bm h_u$. We use independent priors for the $\theta_{t\ell}$'s and $u_{t\ell}$'s and, following \citet{chipman2010bart}, set $\theta_{t\ell} \iid \Normal(0, \sigma^2_\theta)$.

For the sake of computational convenience, we do not use \eqref{eq:integrated-multi} directly, but instead augment the data with latent variables $Z_i \indep \Normal\{\theta_0 + h_\theta(\bX_i), 1\}$ such that $Y_i > 0$ if-and-only-if $Z_i > 0$ \citep{albert1993bayesian}. Before computing \eqref{eq:integrated-multi} we first sample the $Z_i$'s from a $\Normal\{\theta_0 + h_\theta(\bX_i), 1\}$ distribution, truncated to $(-\infty,0)$ or $(0,\infty)$ according as $Y_i = 0$ or $Y_i > 0$. We then compute the integrated likelihood 
\begin{align}
  \label{eq:probit-marginal}
  \begin{split}
  \Lambda(\Tree_t)
  &=
  \pi_\Tree(\Tree_t) \,
  \prod_{\ell \in \Leaves_t} 
  \left[ \int \prod_{i : \bX_i \leadsto (t,\ell)} \Normal\{Z_i \mid \theta_0 + h_\theta(\bX_i), 1\} \  \Normal(\theta_{t\ell} \mid 0, \sigma^2_\theta) \ d\theta_{t\ell}\right. \\
  &\qquad \times \left. \int \prod_{i: Z_i > 0, \bX_i \leadsto (t,\ell)} f\{Y_i \mid \bm h_u(\bX_i), \bomega\} \ \pi_u(u_{t\ell}) \ du_{t\ell}  \right] \\
  &= \pi_\Tree(\Tree_t) \prod_{\ell \in \Leaves_t} L_\theta(t,\ell) \cdot L_u(t,\ell)\ .
\end{split}
\end{align}
Notice that $L_\theta(t,\ell)$ does not depend on our choice for the distribution of the non-zero $Y_i$'s and can be computed in closed form; an expression for $L_\theta(t,\ell)$ is given in the web appendix. Hence, all that must be done to apply the probit-based hurdle model is to be able to compute $L_u(t,\ell)$ in closed form.

\subsection{Gamma hurdle models}

\label{sec:gamma}

Our semiparametric gamma hurdle model sets $Y_i \sim \Gam[\alpha, \alpha \exp\{\lambda_0 + h_\lambda(\bx)\}]$ conditional on $Y_i > 0$ and $\bX_i = \bx$,
where $\Gam(\alpha,\beta)$ is parameterized to have mean $\alpha / \beta$ and variance $\alpha/\beta^2$. We model $h_\theta(\bx)$ and $h_\lambda(\bx)$ with a shared forest, 
\begin{math}
  h_\theta(\bx) = \sum_{t = 1}^T g(\bx ; \Tree_t, \sM_{\theta,t}),
\end{math}
and
\begin{math}
  h_\lambda(\bx) = \sum_{t = 1}^T g(\bx ; \Tree_t, \sM_{\lambda,t}).
\end{math}
Note that, under this model, we have
\begin{align}
  \label{eq:gamma-moments}
  \begin{split}
  E(Y_i \mid Y_i > 0, \bm X_i = \bx, \bm h, \alpha)
  &=
  \exp\left\{ -\lambda_0 - h_\lambda(\bx) \right\},
  \\
  \Var(Y_i \mid Y_i > 0, \bX_i = \bx, \bm h, \alpha)
  &=
    \frac{\exp [-2 \{\lambda_0 + h_\lambda (\bx)\}]}{\alpha}
    ,
  \end{split}
\end{align}
so that the conditional standard deviation of $Y_i$ is proportional to its mean.

The leaf-specific parameters for $h_\lambda(\bx)$ are given log-gamma priors $\lambda_{t\ell} \sim \log \Gam(\alpha_\lambda, \beta_\lambda)$. The log-gamma prior is chosen because it is conjugate to the gamma likelihood and makes computation of \eqref{eq:probit-marginal} tractable. Under this prior for the leaf parameters, the gamma hurdle model is immediately applicable provided that we can compute the likelihood factor
\begin{align*}
  L_\lambda(t,\ell)
  =
  \int \prod_{i \in \ell: Y_i > 0}
  \Gam[Y_i \mid \alpha, \alpha \exp\left\{  \lambda_0 + h_\lambda(\bx)  \right\}]
  \,
  \log \Gam\{\lambda_{t\ell} \mid \alpha_\lambda, \beta_\lambda\}.
\end{align*}
To do this, similar to \citet{murray2017log} for loglinear models, we define
\begin{math}
  \eta_i
  =
  \alpha
  \exp   \{\lambda_0 + h_\lambda(\bX_i) - g(\bX_i ; \Tree_t, \sM_{\lambda,t})\}.
\end{math}
By analogy with the usual Bayesian backfitting algorithm of \citet{chipman2010bart}, the $\eta_i$'s play the role of the backfitted response. Let $A_\ell = \{i : i \in \ell, Y_i > 0\}$ and $N_\ell = |A_\ell|$. Then 
\begin{align*}
  \prod_{i \in A_\ell}
  \Gam[Y_i \mid \alpha,
       \alpha \exp\left\{ \lambda_0 + h_\lambda(\bX_i) \right\}]
  =
  \left(\prod_{i \in A_\ell} \frac{(Y_i \eta_i)^\alpha}{Y_i \Gamma(\alpha)}\right)
  \exp\left(  \alpha N_{\ell} \lambda_{t\ell} - \sum_{i \in A_\ell} Y_i \eta_i e^{\lambda_{t\ell}}  \right).
\end{align*}
Integrating against the $\log\Gam(\lambda_{t\ell} \mid \alpha_\lambda, \beta_\lambda)$ density gives
\begin{align*}
  L_\lambda(t,\ell)
  =
  \frac{\prod_{i \in A_\ell} \eta_i^\alpha Y_i^{\alpha-1}}{\Gamma(\alpha)^{N_\ell}}
  \cdot
  \frac{\beta_\lambda^{\alpha_\lambda}}{\Gamma(\alpha_\lambda)}
  \cdot
  \frac{\Gamma(\alpha_\lambda + N_\ell \alpha)}
  {(\beta_\lambda + \sum_{i \in A_\ell} Y_i \eta_i)^{\alpha_\lambda + N_\ell \alpha}}.
\end{align*}
Hence \eqref{eq:probit-marginal} can be computed in closed form. Additionally, by conjugacy of the log-gamma distribution, we have the full conditionals
\begin{align}
  \label{eq:lambda-full-conditional}
  \lambda_{t\ell} \sim \log \Gam\left(\alpha_\lambda + \alpha N_\ell, \beta_\lambda + \sum_{i \in A_\ell} Y_i \eta_i\right).
\end{align}
A detailed Markov chain Monte Carlo algorithm is given in the web appendix.

\subsection{Log-normal hurdle model}

\label{sec:lognormal}

A shortcoming of the gamma hurdle model is that the relationship between $\bx$ and the variance is captured entirely through the mean. As an alternative, we propose the heteroskedastic log-normal hurdle model with $\pi(\bx) = \Phi\{\theta_0 + h(\bx)\}$ and $(Y_i \mid Y_i > 0, \bX_i = \bx, h, mu, \sigma^2) \sim \Lognormal\{\mu(\bx), \sigma^2(\bx)\}$.
We again use a shared forest to model the three functions
\begin{alignat*}{3}
  \mu(\bx) &= \sum_{t = 1}^T g(\bx ; \Tree_t, \sM_{\mu,t}), &\qquad&&
  \sigma^{-2}(\bx) &= \exp\left\{ \lambda_0 + \sum_{t=1}^T g(\bx ; \Tree_t, \sM_{\lambda,t}) \right\}, \\
  h(\bx) &= \sum_{t=1}^T g(\bx ; \Tree_t, \sM_{\theta,t}).
\end{alignat*}
The resulting model for the mean and variance of $(Y_i \mid Y_i > 0, \bX_i = \bx)$ is given by
$m(\bx) = \exp\{\mu(\bx) + \sigma^2(\bx) / 2\}$ and $s^2(\bx) = m(\bx)^2 [\exp\{\sigma^2(\bx)\} - 1]$.

Like the gamma hurdle model, when a \emph{homoskedastic} model for $\log Y_i$ is used, we find that the mean $m(\bx)$ is proportional to the standard deviation $s(\bx)$. By modeling $\sigma^2(\bx)$ nonparametrically, however, we allow for more complex relationships between $m(\bx)$ and $s(\bx)$. Our heteroskedastic model for the $\log Y_i$'s is similar to the heteroskedastic BART models proposed by \citet{murray2017log} and \citet{pratola2017heteroscedastic}, but our model differs in two respects. First, the trees used to model the mean and variance functions are shared, which is helpful because the variance function $\sigma^2(\bx)$ is generally much more weakly identified than the mean $\mu(\bx)$. Second, our choice of prior for $\sigma^2(\bx)$ will explicitly shrink the posterior model towards a homoskedastic model; see Section~\ref{sec:prior-lognormal}.

Let $\mu_{t\ell}$ and $\lambda_{t\ell}$ be the leaf parameters associated to leaf $\ell$ of $\Tree_t$ for $\mu(\cdot)$ and $\sigma(\cdot)$ respectively and let $\tau_{t\ell} = \exp(\lambda_{t\ell})$. We use a normal-gamma prior for $(\mu_{t\ell}, \tau_{t\ell})$, i.e., 
\begin{math}
  \tau_{t\ell} \sim \Gam(\alpha_\lambda, \beta_\lambda)
\end{math}
and
\begin{math}
  \mu_{t\ell} \sim \Normal\{0, 1/(\kappa \tau_{t\ell})\}.
\end{math}
This normal-gamma prior allows for computation of the likelihood factor
\begin{align*}
  L_{\mu,\lambda}(t,\ell)
  &=
  \int \prod_{i: \bX_i \leadsto (t,\ell), Y_i > 0}
  \log \Normal\{Y_i \mid \mu(\bX_i), \sigma^2(\bX_i)\}
  \\
  &\qquad\times 
  \Normal\{\mu_{t\ell} \mid 0, 1/(\kappa \tau_{t\ell})\}
  \
  \Gam(\tau_{t\ell} \mid \alpha_\lambda, \beta_\lambda)
  \
  d\mu_{t\ell} \ d\tau_{t\ell}.
\end{align*}
Let $W_i = \log Y_i$ and suppose $\bX_i \leadsto (t,\ell)$. Then, conditional on $Y_i > 0$, we have
\begin{math}
  W_i \sim \Normal(\eta_i + \mu_{t\ell} , \frac{1}{\nu_i \tau_{t\ell}}),
\end{math}
where 
\begin{math}
  \eta_i = \sum_{j \ne t} g(\bX_i; \Tree_j, \sM_{\mu,j}),
\end{math}
and
\begin{math}
  \nu_i = \exp\left\{  \lambda_0 + \sum_{j \ne t}  g(\bX_i; \Tree_j, \sM_{\lambda,j})\right\}.
\end{math}
Let $Q_i = W_i - \eta_i$ and $A(\ell) = \{i : \bX_i \leadsto (t,\ell), Y_i > 0\}$; $Q_i$ and $\nu_i$ are analogous to the backfitted response in the usual Bayesian backfitting algorithm. We have
\begin{align*}
  \prod_{i \in A(\ell)} \Normal\left(W_i \mid \eta_i + \mu_{t\ell}, \frac 1 {\nu_i \tau_{t\ell}}\right)
  =
  \prod_{i \in A(\ell)}
  \left( \frac{\nu_i \tau_{t\ell}}{2\pi} \right)^{1/2}
  \exp\left\{ -\frac{\nu_i \tau_{t\ell}}{2} (Q_i - \mu_{t\ell})^2 \right\} .
\end{align*}
This likelihood is conjugate to the normal-gamma prior for $(\mu_{t\ell}, \tau_{t\ell})$, and routine computations give the expression
\begin{align*}
  \left( \prod_i \sqrt{\frac{\nu_i}{2\pi}} \right)
  \sqrt{\frac{\kappa}{\kappa + w_\ell}} 
  \frac{\beta_\lambda^{\alpha_\lambda} \Gamma(\alpha_\lambda + N_\ell / 2)}{\Gamma(\alpha_\lambda)}
  \left( \beta_\lambda + \frac{S^2_{\ell}}{2} + \frac{\kappa w_\ell \bar Q_\ell^2}{2(\kappa + w_\ell)} \right)^{-(\alpha_\lambda + N_{\ell}/2)},
\end{align*}
for $L_{\mu,\lambda}(t,\ell)$ where 
\begin{align*}
  \bar Q_\ell = \frac{\sum_{i \in A(\ell)} \nu_i Q_i}{\sum_{i \in A(\ell)} \nu_i},
  \qquad
  w_\ell = \sum_{i \in A(\ell)} \nu_i,
  \qquad \text{and} \qquad
  S^2_\ell = \sum_{i \in A(\ell)} \nu_i (Q_i - \bar Q_\ell)^2.
\end{align*}
We again have a closed form for \eqref{eq:probit-marginal}. Moreover, we also have the following full conditionals for the leaf parameters:
\begin{align}
  \label{eq:ng-full-conditional}
  \tau_{t\ell} \sim \Gam(\widehat \alpha_\ell, \widehat \beta_\ell),
  \qquad
  \text{and}
  \qquad
  \mu_{t\ell} \sim \Normal\{\widehat \mu_\ell, 1/(\widehat \kappa_\ell \tau_{t\ell})\},
\end{align}
where
\begin{alignat*}{3}
  \widehat \alpha_\ell &= \alpha_\lambda + N_\ell / 2,
  &&\qquad&
  \widehat \beta_\ell &= \beta_\lambda + \frac{S^2_\ell}{2} + \frac{\bar Q_\ell^2 \kappa w_\ell}{2(\kappa + w_\ell)},
  \\
  \widehat \kappa_\ell &= \kappa + w_\ell,
  &&\qquad&
  \widehat \mu_\ell &= \frac{\sum_{i \in A(\ell)} v_i Q_i}{\widehat \kappa}.
\end{alignat*}
Additional details for the various steps of the MCMC algorithm are deferred to the web appendix.

\subsection{Prior specification}

\label{sec:prior-lognormal}

An advantage of the BART framework is that there exist standard ``default'' priors which have proven to work remarkably well in practice. In particular, very little tuning is required to obtain an acceptable baseline level of performance. We develop default priors for the gamma hurdle and log-normal hurdle models we consider here. For both models, we will use the default prior recommended by \citet{chipman2010bart} for the $\theta_{t\ell}$'s. Additionally, we apply a quantile normalization separately to each column of the design matrix $\bX$ so that the predictors are distributed approximately uniformly on $[0,1]$. 

We first give a prior specification for the log-normal hurdle model. As a preprocessing step, we work with $W_i = \log Y_i$; further, we standardize the finite $W_i$'s to have mean $0$ and standard error $1$. In order for the prior to be stable as the number of trees is increased, we choose the hyperparameters so that $E(\lambda_{t\ell}) = 0$ and $\Var(\lambda_{t\ell}) = a_\lambda^2 / T$, and similarly for $\mu_{t\ell}$ and $\theta_{t\ell}$. This ensures that the stochastic process $\sum_{t=1}^T g(\bx; \Tree_t, \sM_t)$ converges to a Gaussian process as $T \to \infty$ so that the prior is stable under adding additional trees.

Appropriate values for $(\alpha_\lambda, \beta_\lambda)$ can be obtained by solving the equations
\begin{align}
  \label{eq:mean}
  E(\lambda_{t\ell}) &= \psi(\alpha_\lambda) - \log \beta_\lambda = 0, \\
  \label{eq:var}
  \Var(\lambda_{t\ell}) &= \psi'(\alpha_\lambda) = a^2_\lambda / T.
\end{align}
Noting that $\psi'(\alpha) \approx \alpha^{-1}$, \eqref{eq:var} implies that for moderate values of $T$ we will have $\alpha \approx T/a_{\lambda}^2$. Additionally, noting that $\psi(\alpha) \approx \log(\alpha)$, \eqref{eq:mean} implies that $\alpha_\lambda \approx \beta_{\lambda}$; in particular, both $\alpha_\lambda$ and $\beta_{\lambda}$ are roughly proportional to $T$. 

As there is typically less information in the data about the second order effect $\sigma^2(\bx)$ than the first order effect $\mu(\bx)$, it is sensible to shrink our model towards a homoskedastic model. Note that if all the $\lambda_{t\ell}$'s are equal to $0$ then the variance function reduces to $\sigma^{2}(\bx) = \exp(-\lambda_0)$ so that the model is homoskedastic. Accordingly, we place a $\Halfcauchy(0,1)$ on the baseline standard deviation $\sigma_0 = \exp(-\lambda_0 / 2)$ and shrink the $\lambda_{t\ell}$'s heavily to zero. 
As a default, we have found $a_\lambda = 0.5$ to work well in practice; alternatively, one might set $a_\lambda \sim \Halfcauchy(0,1)$ to allow the model to adaptively determine the amount of heteroskedasticity in the data.

Next, by analogy with the prior specification of \citet{chipman2010bart}, we ensure that the $\mu_{t\ell}$'s marginally have mean 0 and standard deviation $3/(k_\mu \sqrt{T})$ by noting that
\begin{math}
  \Var(\mu_{t\ell}) = \beta / \{(\alpha - 1)\kappa\}.
\end{math}
As noted above, for moderate $T$ we will have $\alpha_{\lambda} \approx \beta_{\lambda} \propto T$, so that $\Var(\mu_{t\ell}) \approx \kappa^{-1}$. This suggests setting $\kappa^{-1/2} = k_{\mu} / \sqrt{T}$ (or giving $\kappa^{-1/2}$ a prior centered at this value). Here $k_\mu$ is a tuning parameter which controls the signal-to-noise ratio and by default we set $k_{\mu} = 1.5$.

We recommend a similar default prior for the gamma model. We first scale the non-zero $Y_i$'s to have mean $1$. As before, we impose the restrictions $E(\lambda_{t\ell}) = 0$ and $\Var(\lambda_{t\ell}) = a_\lambda^2 / T$ so that $h_{\lambda}(\bx) \asim \Normal(0, a_\lambda^2)$. This can be accomplished by solving the system of equations~(\ref{eq:mean}, \ref{eq:var}). As a default, we set $a_\lambda = k_{\lambda}\sqrt{\Var(\log Y_i \mid Y_i > 0)}$ where $k_\lambda$ is a user-specified tuning parameter which we set to $1.5$. Additionally, we require a prior for the shape parameter $\alpha$. From \eqref{eq:gamma-moments} we see that $1/\alpha$ is a dispersion parameter. We use a weakly informative half-Cauchy prior $\alpha^{-1/2} \sim \Halfcauchy(0,A)$ for some $A > 0$. For the MEPS data in particular we set $A = 1$ to encourage small values of $\alpha$, as medical expenditures are highly right-skewed.

\subsection{Identifiability of the model components}

Given that the hurdle models we have proposed are mixture models, there is a question of whether the models we have defined here are identifiable. Let $\mathcal X$ denote a predictor space and $(\mathcal Y, \mathscr B)$ be a measurable space. Given a class $\{F_{\theta} : \theta \in \Theta\}$ where $F_{\theta} : \mathscr B \times \mathcal X \to [0,1]$ is a probability distribution on $(\mathcal Y, \mathscr B)$ for every $\bx \in \mathcal X$, the parameter $\theta$ is called \emph{identifiable} if the mapping $\theta \mapsto F_{\theta}$ is one-to-one (\citealp{lehmann2006theory}, Definition 1.5.2). General forms of the hurdle model may not be identifiable, particularly when we are mixing a point mass at $0$ with a distribution that also is supported at $0$, or if the positive part has probability $0$. The following lemma shows that this is essentially the only case in which we might run into identifiability issues. A proof this result is given in the web appendix.

\begin{lemma}
  \label{prop:ident}
  Let $\mathcal X$ denote an arbitrary set, $(\mathbb R, \mathscr B)$ the Borel measurable reals, and $\delta_0$ the point-mass distribution at $0$.
  Let $\mathscr G$ denote the set of conditional distributions with no atoms at $0$
  \begin{align*}
    \mathscr{G}= & \begin{Bmatrix}G:\mathscr{B}\times\mathcal{X}\to[0,1] & \text{such that } & G_{\bx}(\cdot)\text{ is a probability distribution}\\
      & \text{and} & G_{\bx}(\{0\})=0\text{ for all }\bm{x}\in\mathcal{X}
    \end{Bmatrix}
  \end{align*}
  and let $\mathscr P$ be the collection of conditional probabilities which are bounded by $1$, 
  \begin{math}
    \mathscr P
    =
    \left\{
    \pi: \mathcal X \to [0,1)
    \right\}.
  \end{math}
  Let $\mathscr M$ denote the collection of conditional distributions on $(\mathbb R, \mathscr B)$ which are not identically $0$, 
  \begin{align*}
    \mathscr{M}= & \begin{Bmatrix}M:\mathscr{B}\times\mathcal{X}\to[0,1] & \text{such that } & F_{\bx}(\cdot)\text{ is a probability distribution}\\
      & \text{and} & F_{\bx}(\{0\}) \ne 1\text{ for all }\bm{x}\in\mathcal{X}
    \end{Bmatrix}.
  \end{align*}
  Then the mapping $\mathscr G \times \mathscr P \to \mathscr M$ given by
  \begin{math}
    (G, \pi) \mapsto \pi(\bm x) \delta_0 + \{1-\pi(\bm x)\} G_{\bx},
  \end{math}
  is a bijection. 
\end{lemma}

A consequence of this result is that the semiparametric hurdle models developed in Section~\ref{sec:gamma} and Section~\ref{sec:lognormal} are also identifiable when the model parameters are understood to be the nonparametrically-specified functions $\bm h$ and the parametric component $\bomega$ (noting that $(\bm h, \bomega)$ maps in a one-to-one fashion to $(G,\pi)$).
The individual trees in the ensemble are, however, not identifiable, as the collection of possible regression trees is an overcomplete basis. In practice, we are usually only interested in recovering $\bm h$ rather than the individual trees, so that this lack of identifiability is not a concern.

\section{Simulation study}

\label{sec:simulation}

In this section, we examine the benefits of sharing information across related tasks using a simple simulation study. We consider a mixed response
\begin{align}
  \label{eq:sim-model}
  \Pr(Z_i = 1 \mid \bX_i = \bx)
  = \Phi\{\sigma_\theta \, h(\bx)\}, \qquad (Y_i \mid \bX_i = \bx) \sim \Normal\{h(\bx), \sigma^2\},
\end{align}
with $(Z_i \perp Y_i \mid \bX_i = \bx)$. This is similar to the zero-inflated response setting, but with the continuous portion of the distribution always observed (see also Example~\ref{ex:mixed}). Note that the information in $\bX_i$ is captured by the one-dimensional summary $h(\bX_i)$ which is shared across both models. We emphasize that the structure \eqref{eq:sim-model} is not assumed by the shared forest model - only the basis functions are shared - and must effectively be learned from the data. We consider the benchmark function given by \citet{friedman1991multivariate}
\begin{align*}
  h(\bx) = 10 \sin(\pi x_1 x_2) + 20 (x_3 - 0.5)^2 + 10 x_4 + 5 x_5.
\end{align*}
We sample $\bX_i$ uniformly distributed on $[0,1]^P$; if $P > 5$ then the predictors $X_{i6}, \ldots, X_{iP}$ have no influence on the response. We compare the shared forest to an approach which fits separate BART models to $(Z_i \mid \bX_i = \bx)$ and $(Y_i \mid \bX_i = \bx)$ so that information is not shared across tasks. Our focus is on how well these models estimate $\Pr(Z_i = 1 \mid \bX_i) = \pi(\bX_i)$ as measured by the cross-entropy between the true and estimated $\pi$'s,
\begin{align*}
  \text{Loss}
  = \int \left[\pi(\bx) \log \left( \frac{\pi(\bx)}{\widehat \pi(\bx)} \right)
  + \{1 - \pi(\bx)\} \log \left( \frac{1 - \pi(\bx)}{1 - \widehat \pi(\bx)} \right)
    \right]
  \ d\bx,
\end{align*}
which is computed by Monte Carlo integration. We focus on the setting in which the continuous response $Y_i$ is relatively informative while the information contained in $Z_i$ is relatively weak by fixing $\sigma^2 = 1$. We consider a training set of size $n = 250$ for both the $Y_i$'s and the $Z_i$'s.

\begin{figure}
  \centering
  \includegraphics[width = 1\textwidth]{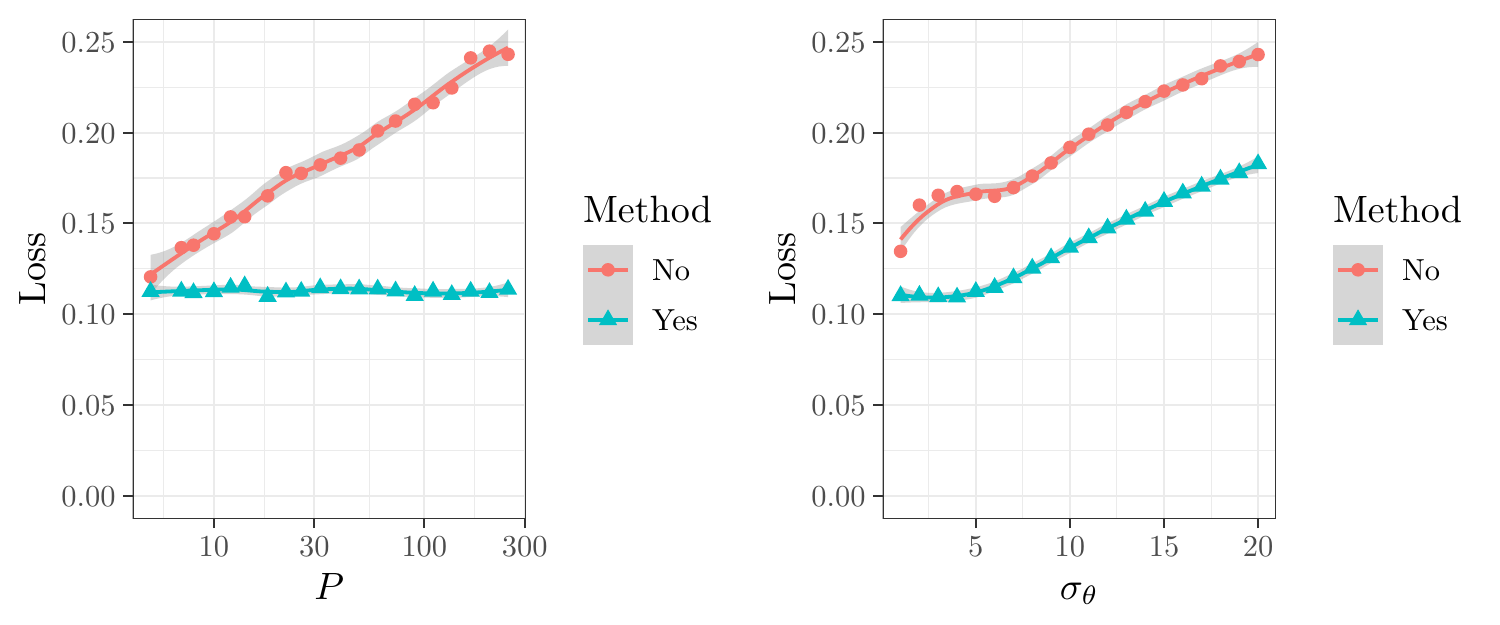}
  \caption{Left: average value of Loss for $\log P \in (\log 5, \log 250)$ averaged over 20 replications. Right: average value of Loss for $\sigma_\theta$ ranging from $1$ to $20$. ``No'' indicates the single BART model while ``Yes'' indicates use of the shared forest. This figure appears in color in the electronic version.}
  \label{fig:simulation-share-noshare}
\end{figure}

Results are given in Figure~\ref{fig:simulation-share-noshare}, with $20$ replications per simulation setting. In the left panel, we fix $\sigma_\theta= 4$ (roughly corresponding to $\pi(\bX_i) \sim \Uniform(0,1)$) and examine how sharing impacts the loss as $P$ varies from $P = 5$ to $P = 250$. We see that, as the variable selection task becomes more difficult, the model which does not share information is far more sensitive to irrelevant predictors than the model which does share. This is because the $Y_i$'s are much more informative about the relevant predictors than the $Z_i$'s, so that the shared model can do a much better job of selecting the relevant predictors. In the right panel, we fix $P = 20$ and vary the signal level $\sigma_\theta$ from $1$ to $20$. In this case, the gain from sharing is essentially constant, with higher losses for higher signal levels.

\begin{figure}
  \centering
  \includegraphics[width=.7\textwidth]{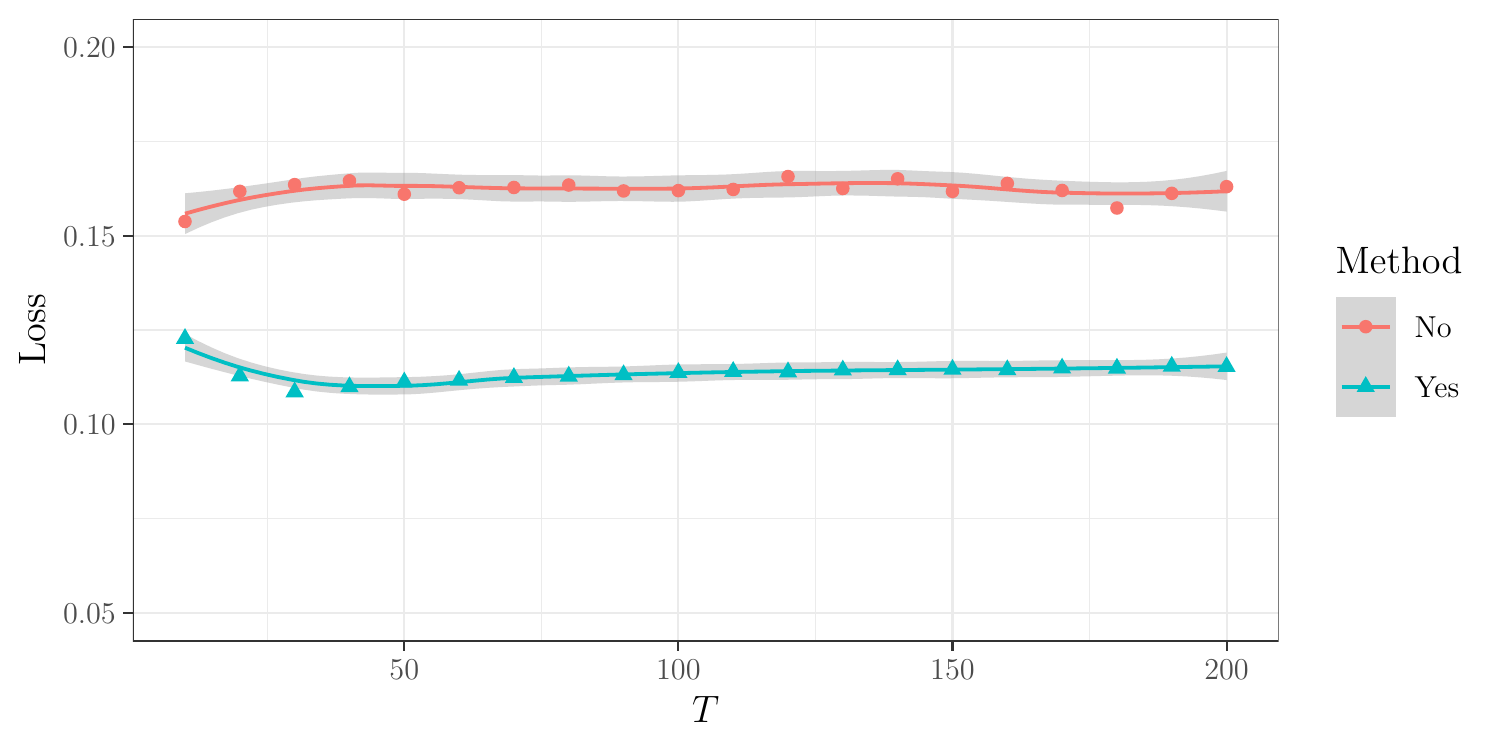}
  \caption{Average values of Loss for $T$ evenly space in $[10, 200]$ averaged over 20 replications.  This figure appears in color in the electronic version.}
  \label{fig:tree-sim}
\end{figure}

A potentially important tuning parameter in BART models is the number of trees $T$ used in the ensemble. Other works have supported the following overall trend: predictive performance of BART models is typically insensitive to the number of trees included, provided that we include sufficiently many. We find that this behavior holds for the shared forests model as well, with Figure~\ref{fig:tree-sim} summarizing the results of the simulation experiment with $P = 20$ and $\sigma_\theta = 4$ fixed as a function of $T$. As before, results are based on $20$ replications of the experiment for each setting. As expected we see a slight decrease in performance as $T$ increases from the optimal choice.

\section{Analysis of MEPS data}

\label{sec:analysis}

Our motivating example is from the 2015 Medical Expenditure Panel Survey (MEPS).  The MEPS study \citep{natarajan2008variance} was developed to estimate national and regional health care use and expenditures in the United States. We first illustrate the capability of the proposed model to effectively capture heteroskedasticity in the MEPS data. We analyze data from 10,729 adult females who participated in the survey, focusing on the total medical expenditure during 2015, denoted $Y_i$. Previous analyses of this dataset have suggested taking $\Var(Y_i) = \phi E(Y_i)^{1.5}$ \citep{blough2000using, natarajan2008variance}. We consider a list of predictors including, among other things, age, race, family income, whether the individual smokes, perceived health, body mass index, and number of visits to the dentist over the survey period; a full list of predictors is given in the web appendix.

We fit the log-normal hurdle and gamma hurdle models to the data. We examine the fit of these models to the positive part of the data $(Y_i \mid Y_i > 0, \bX = \bx)$ by considering the generalized residuals \citep{cox1968general} $r_i = \Phi^{-1}\{\widehat F_{\bX_i}(Y_i)\}$ where $\widehat F_{\bx}$ is an estimate of the cdf of $(Y_i \mid Y_i > 0, \bX_i = \bx)$ obtained from the model. In the case of the log-normal hurdle model, $r_i$ is equivalent to the usual standardized residual $\{\log Y_i - \widehat \mu_i(\bx)\} / \widehat \sigma(\bx)$; for comparison, we also consider the raw residuals $\log Y_i - \widehat \mu_i(\bx)$ to examine the effect of heteroskedasticity on the model fit.

\begin{figure}
  \centering
  \includegraphics[width=.9\textwidth]{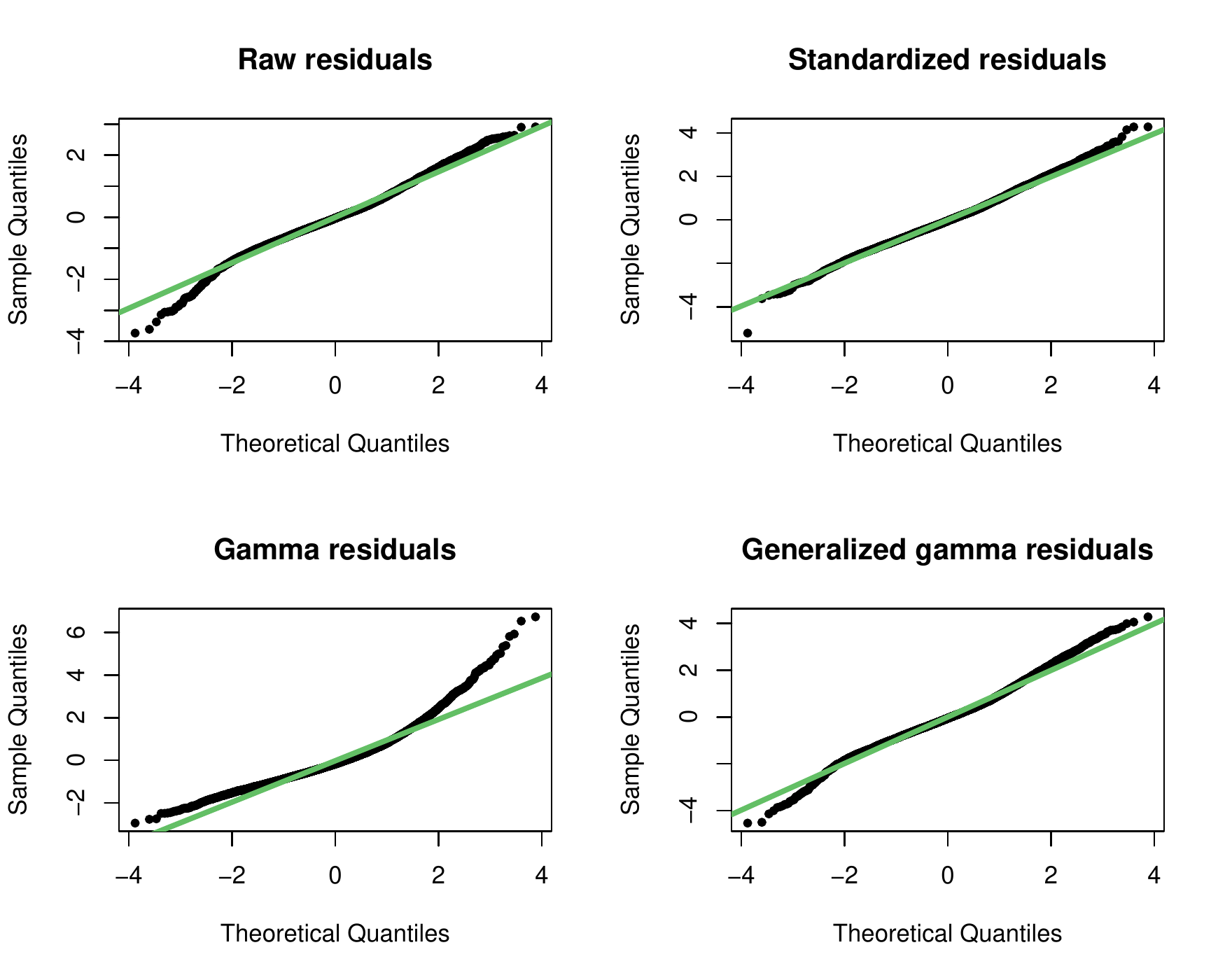}
  \caption{Quantile-quantile plots comparing the residuals $r_i$ for each model to a reference normal distribution. The top panels give the raw residuals $\log Y_i - \widehat \mu(\bX_i)$ (left) and standardized residual $r_i$ for the log-normal hurdle model (right). The bottom panels give the residuals $r_i$ for the gamma hurdle (left) and the generalized gamma hurdle (right) models. This figure appears in color in the electronic version.}
  \label{fig:qq}
\end{figure}
Quantile-quantile plots of the residuals compared to a reference Gaussian distribution are given in Figure~\ref{fig:qq}. We see that the log-normal hurdle model fits the data very well. Additionally, we see that ignoring heteroskedasticity causes a poor fit in the left tail of the data, corresponding to individuals with lower healthcare costs. By comparison, the gamma model fits poorly. We also consider a generalized gamma distribution \citep{stacy1962generalization} which models $Y_i^\phi$ with a gamma distribution, where $\phi$ is learned from the data. The generalized gamma model fits roughly as well as a homoskedastic log-normal model, but is inferior to the heteroskedastic log-normal model due to the stringent relationship between the mean and the variance implied by the generalized gamma model.

\begin{figure}[!ht]
  \centering
  \includegraphics[width=.9\textwidth]{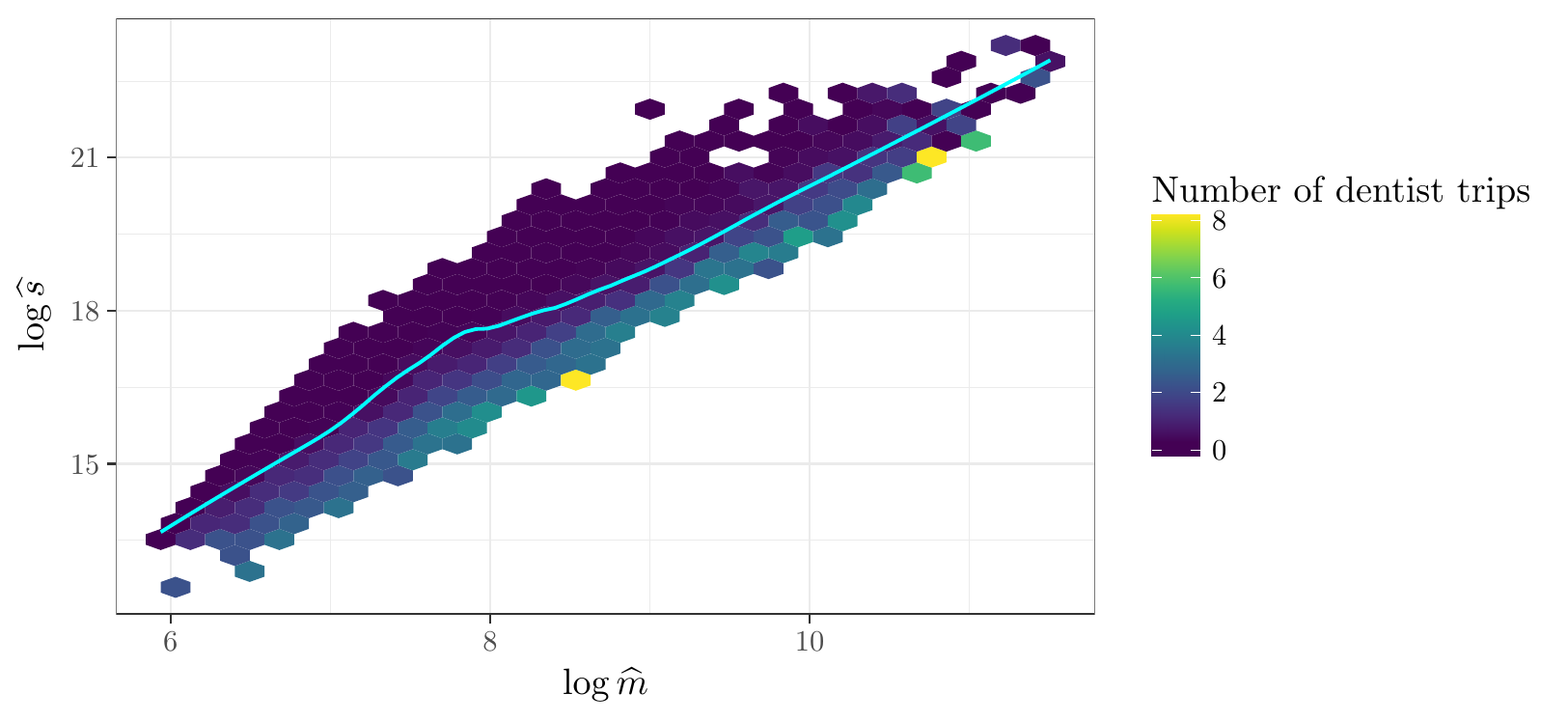}
  \includegraphics[width=.9\textwidth]{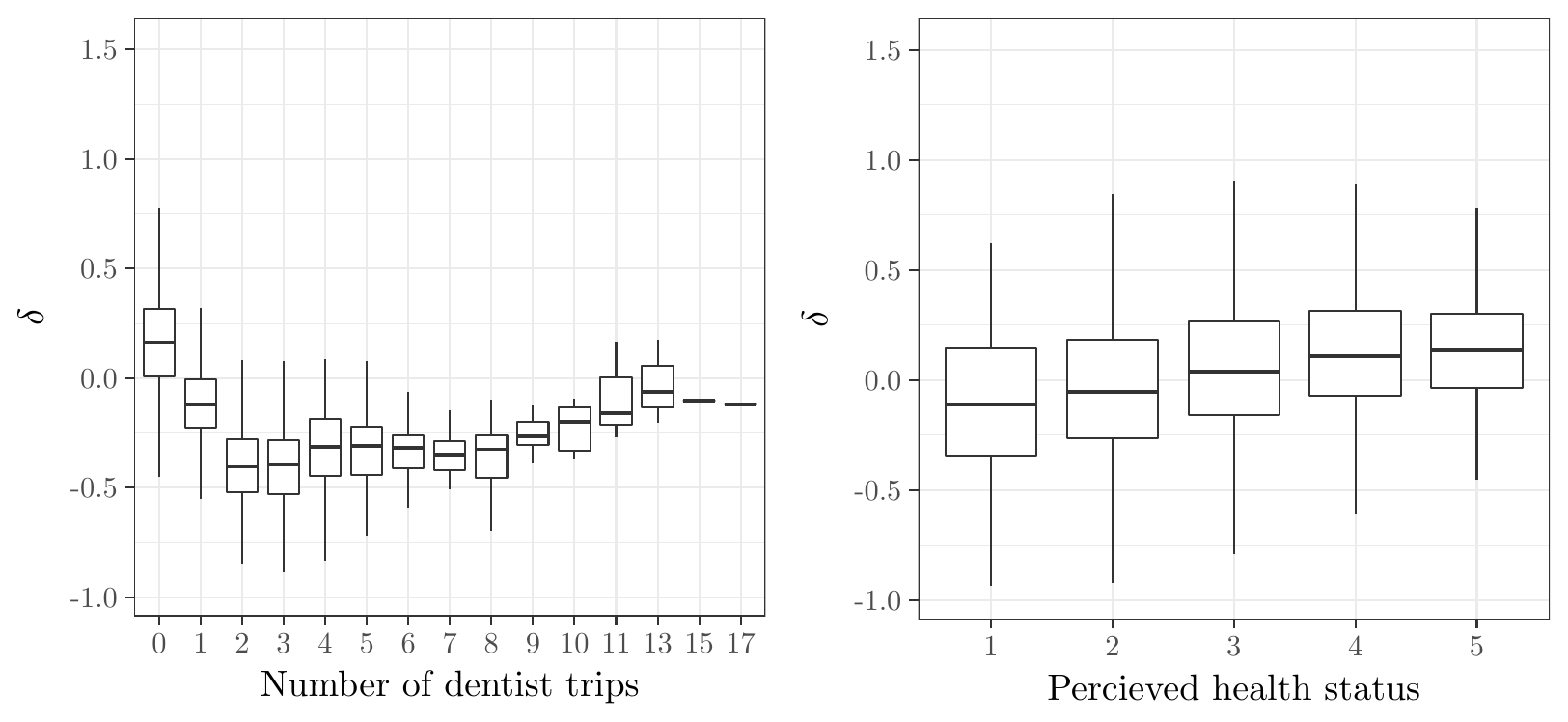}
  \caption{Top: Plot of $\widehat s^2(\bm X_i)$ against $\widehat m(\bX_i)$ on the log-log scale; individual points are binned into hexagons, which are shaded according to the number of dentist visits the subject has. Bottom: boxplots of $\delta$ for the number of dentist trips and perceived health status. This figure appears in color in the electronic version.}
  \label{fig:hetero}
\end{figure}

In addition to fitting the data well, the heteroskedastic log-normal model provides several interesting insights into the nature of the heteroskedasticity in the data. Let $\widehat m(\bx)$ and $\widehat s(\bx)$ denote the posterior mean of $m(\bx)$ and $s(\bx)$ given in Section~\ref{sec:lognormal}. The top panel of Figure~\ref{fig:hetero} gives a plot of $\widehat m(\bX_i)$ against $\widehat s(\bX_i)$ on the log-log scale. To aide visualization, points with similar values of $(\widehat m(\bX_i), \widehat s(\bX_i))$ are grouped into hexagonal tiles and are shaded according to the \emph{average number of dentist visits per subject} within each tile.

There are several interesting features of the top panel of Figure~\ref{fig:hetero}. First, there is near-linear relationship between $\log \widehat m(\bX_i)$ and $\log \widehat s(\bX_i)$. An ordinary least squares (OLS) fit of $\log \widehat m(\bX_i)$ to $\log \widehat s(\bX_i)$ has slope $0.7556$ and an $R^2$ of 82\%. Hence, the OLS fit suggests the approximation $\widehat s^2(\bX_i) \propto \widehat m(\bX_i)^{1.511}$, which agrees nearly exactly with \citet{blough2000using}. Second, by shading the hexagonal tiles by the number of dentist visits, we see clearly that the mean \emph{does not} account for all of the heteroskedasticity due to the predictors. We see, for example, that individuals with lower numbers of visits to the dentist tend to have a standard deviation which is higher than what would be predicted by the mean alone. To understand this relationship better, we let
\begin{math}
\delta = \log \widehat s(\bX_i) - 0.7556 \log \widehat m(\bX_i) - 2.672
\end{math}
denote the residual in predicting $\log \widehat s(\bX_i)$ with $\log \widehat m(\bX_i)$ by OLS. The bottom panel of Figure~\ref{fig:hetero} shows how the distribution of $\delta$ varies across the number of dentist trips and the individual's perceived health status. We see first that individuals with fewer dentist trips have standard deviations which are larger than what would be predicted using only the mean; similarly, individuals with higher perceived health status scores (corresponding to \emph{lower} perceived health) also tend to have higher variability than would be predicted by the mean alone.

To assess whether there is a benefit of using the shared forests methodology for the MEPS data, we compute the log pseudo-marginal likelihood for the shared forest model and an equivalent model which does not share the trees across model components. Specifically, we fit the heteroskedastic BART (HBART) model of \citet{pratola2017heteroscedastic}, which sets $Y_i = g(\bX_i) + s(X_i) \epsilon_i$ with $g(\bX_i)$ and $\log s(\bX_i)$ given BART priors. HBART was fit to the non-zero observed $Y_i$'s. The probability of a zero response was modeled using a binary BART model with a probit link, i.e., $\Pr(Y_i = 0 \mid \bX_i = \bx) = \Phi\{h(\bx)\}$ where $h(\cdot)$ was given a BART prior. The log pseudo-marginal likelihood (LPML) is given by $\LPML = \sum_{i = 1}^n \log \CPO_i$ where $\CPO_i = f(Y_i \mid \bY_{-i}, \bX)$ is the predictive density of the $i^{\text{th}}$ observation given $\bY_{-i} = (Y_1, \ldots, Y_{i-1}, Y_{i+1}, \ldots, Y_n)$ and $\bX = (\bX_1, \ldots, \bX_n)$ \citep{geisser1979predictive}. Results are given in Table~\ref{tab:loo}. The $\LPML$ was computed using the Markov chain output using the \texttt{loo} package in \texttt{R} \citep{vehtari2017practical}. Multiple fits of the model using different seeds for the MCMC algorithm give qualitatively similar results. 

\begin{table}
  \centering
  \begin{tabular}{lrr}
    \toprule 
    & Shared   & Not Shared \\
    \midrule
    Regression & -15166.9 & -15267.2   \\
    Binary     & -1552.8  & -2069.7    \\
    Total      & -16719.7 & -17336.9   \\
    \bottomrule
  \end{tabular}
  \vspace{2em}

  \caption{LPML of the model when the forests are shared across the mean, variance,
    and hurdle components, compared with the LPML when the forests are
    not shared. The row ``Regression'' gives the LPML contribution obtained
    from $[Y_{i}\mid Y_{i}>0]$, while the row ``Binary'' give the LPML
    contribution obtained from $I(Y_{i}>0)$; ``Total'' gives the final
    LPML. }
  \label{tab:loo}
\end{table}

We see that the shared forest gives a substantial boost in $\LPML$ for the binary component of the model. This suggests that the features learned from the continuous part of the model are helpful in determining whether an individual incurs any medical expense. We also observe a less substantial, but still large, improvement in performance for the regression model.

\section{Discussion}

\label{sec:discussion}

In this paper we introduced shared forests and demonstrated their usefulness on both simulated data and data from the MEPS dataset. Additionally, we introduced two novel models for semicontinuous data: a gamma hurdle model and a heteroskedastic log-normal hurdle model. 

There are several promising areas for future work. First, there are other possibilities for sharing information across nonparametric components. Here we have restricted the components to share the same basis function expansion. To make the models more tightly coupled, one might consider shrinking together the coefficients of these expansions; an example where this might be useful is in meta-regression, where one would expect both that features across different studies will exert similar (but not necessarily identical) effects on the outcome. In the other direction, one might allow the models to share a \emph{subset} of the basis functions; for example, each model component might consist of a shared forest combined with an \emph{innovation forest} which is specific to each task. This structure is likely to be useful if only a subset of relevant features are shared across nonparametric components. A special case of such a construction is given by \citet{hahn2017bayesian} to estimate heterogeneous causal effects; in our terminology, their model consists of a shared forest which captures the prognostic features of covariates which are shared across treatment levels $z = 1$ and $z = 0$ and an innovation forest which is specific to the treatment $z = 1$. 

Additionally, \citet{linero2017abayesian} recently demonstrated that the discrete nature of decision trees can lead to suboptimal performance on both a theoretical and practical level, and that this can be corrected by replacing the usual decision trees with \emph{smooth} decision trees. The shared forests framework can easily be extended to allow for smooth decision trees for the homoskedastic log-normal hurdle model, but non-trivial modifications are required to apply this strategy to the heteroskedastic log-normal and gamma hurdle models.

\section{Supporting Information}

A web appendix referenced in Sections 1, 3, and 5, and code implementing the methodology, are available with this paper at the Biometrics website on Wiley Online Library; code is also available at \url{www.github.com/theodds/SharedForestPaper}.
% Code is available at \url{www.github.com/theodds/SharedForestPaper}.

\section*{Acknowledgements}

This work was partially supported by NSF grant DMS-1712870 and by the Office of
the Secretary of Defense, Directorate of Operational Test and Evaluation and the
Test Resource Management Center under the Science of Test research consortium.

\bibliographystyle{apalike}
\bibliography{mybib}

% \label{lastpage}



\section*{Supplementary material}

# Supporting Information for Semiparametric Mixed-Scale Models Using Shared Bayesian Forests

Antonio R. Linero  
Department of Statistics  
Florida State University  
`arlinero@stat.fsu.edu`

Debajyoti Sinha  
Department of Statistics  
Florida State University  
`sinhad@stat.fsu.edu`

Stuart R. Lipsitz  
Department of Medicine  
Brigham and Women's Hospital  
`slipsitz@partners.org`

## S.1   MEPS predictors

The following predictors were used in the analysis of the MEPS dataset:

- `age`: the age of the individual as of 2015.

- `smoke`: whether the individual currently smokes.

- `race`: whether the individual is black, white, Hispanic, Asian, or other.

- `insurance`: whether an individual has private, public, or no health insurance.

- `phealth`: the individuals perceived health (1 to 5).

- `income`: the individual's familial income.

- `meds`: the number of prescription medications the individual is taking.

- `bmi`: the body mass index.

- `education`: the level of completed education.

- `diabetes, stroke, cancer, heart_attack, cognitive_limitations, arthritis`: indicators for whether the subject has suffered from any of these conditions.

- `down`: whether an individual feels down/depressed/hopeless.

- `dentist`: the number of dentist visits over the survey period.

## S.2 MCMC for the log-normal and gamma hurdle models

We provide details for implementing Markov chain Monte Carlo algorithms for the gamma hurdle and log-normal hurdle models. We consider a Markov chain operating on the $\mathcal{T}_t$'s, $\mathcal{M}_t$'s, and non-tree-specific parameters $\boldsymbol{\omega}$. Both approaches use the same basic approach, which is summarized by Algorithm 1.

---
**Algorithm 1** Bayesian backfitting algorithm
---
1: **for** $t = 1, \ldots, T$ **do**
2:      Propose $\mathcal{T}_t' \sim Q(\mathcal{T}_t \to \mathcal{T}_t')$
3:      Sample $U \sim \text{Uniform}(0, 1)$ and compute the acceptance ratio

$$\rho(\mathcal{T}_t, \mathcal{T}_t') = \frac{\Lambda(\mathcal{T}_t')Q(\mathcal{T}_t' \to \mathcal{T}_t)}{\Lambda(\mathcal{T}_t)Q(\mathcal{T}_t \to \mathcal{T}_t')}.$$

4:      If $U \leq \rho(\mathcal{T}_t, \mathcal{T}_t')$, set $\mathcal{T}_t \leftarrow \mathcal{T}_t'$, otherwise leave $\mathcal{T}_t$ unchanged.
5:      Update the leaf node parameters according to (S.1), (7) and/or (8).
6: **end for**
7: Make an update to $\boldsymbol{\omega}$ which leaves its full conditional invariant.
8: **for** $i = 1, \ldots, n$ **do**
9:      Sample $Z_i \sim \text{Normal}(\theta_0 + h_\theta(\boldsymbol{x}), 1)$ truncated to $(-\infty, 0)$ or $(0, \infty)$ according as $Y_i = 0$ or $Y_i > 0$.
10: **end for**
---

To compute $\Lambda(\mathcal{T}_t)$, Algorithm 1 requires the expression

$$L_\theta(t, \ell) = (\sqrt{2\pi})^{-n_\ell} \sqrt{\frac{\sigma_\theta^{-2}}{\sigma_\theta^{-2} + n_\ell}} \exp\left(-\frac{\text{SSE}_\ell}{2} - \frac{n_\ell \sigma_\theta^{-2} \bar{R}_\ell^2}{2(n_\ell + \sigma_\theta^{-2})}\right),$$

where

$$n_\ell = |\{i : \boldsymbol{X}_i \rightsquigarrow (t, \ell)\}|, \qquad \text{SSE}_\ell = \sum_{i:\boldsymbol{X}_i \rightsquigarrow (t,\ell)} (R_i - \bar{R}_\ell)^2,$$

$$\bar{R}_\ell = \frac{1}{n_\ell} \sum_{i:\boldsymbol{X}_i \rightsquigarrow (t,\ell)} R_i, \qquad R_i = Z_i - \theta_0 - \sum_{j \neq t} g(\boldsymbol{X}_i; \mathcal{T}_t, \mathcal{M}_{\theta,t}),$$

which is derived (e.g.) in Kapelner and Bleich (2016). Further, we require the full conditional

$$\theta_{t\ell} \sim \text{Normal}\left(\frac{n_\ell \bar{R}_\ell}{n_\ell + \sigma_\theta^{-2}}, \frac{1}{n_\ell + \sigma_\theta^{-2}}\right). \tag{S.1}$$

For the gamma-hurdle model, $\boldsymbol{\omega}$ consists of just the parameter $\alpha$, which we give the prior $\alpha^{-1/2} \sim \text{Cauchy}_+(0, A)$. Under this prior, the full conditional for $\alpha$ is proportional to

$$\frac{\alpha^{\alpha N}}{\Gamma(\alpha)^N} \left(\prod_{i=1}^n Y_i^\alpha\right) \exp\left[\alpha \sum_i \{\lambda_0 + h_\lambda(\boldsymbol{x})\} - \alpha \sum_i Y_i e^{\lambda_0 + h_\lambda(\boldsymbol{x})}\right] \times \frac{1}{\pi(A\alpha^{3/2} + \alpha^{1/2}/A)},$$

and we update $\alpha$ using slice sampling (Neal, 2003).

For the log-normal hurdle model, $\boldsymbol{\omega}$ consists of the baseline standard deviation $\sigma_0 = \exp(-\lambda_0/2)$. Let $\nu_i = \sigma^2(\boldsymbol{X}_i)/\sigma_0^2$. Then the full conditional of $\sigma_0$ is proportional to

$$\sigma_0^{-N/2} \exp\left[-\frac{1}{\sigma_0^2} \sum_{i=1}^n \nu_i \{Y_i - \mu(\boldsymbol{X}_i)\}^2\right] \frac{1}{\pi(1 + \sigma_0^2)},$$

where $N = |\{i : Y_i > 0\}|$. As before, $\sigma_0$ can be updated via slice sampling.

## S.3 Metropolis-Hastings Algorithm

We construct Metropolis-Hastings the proposal from updating $\mathcal{T}_t$ using the general strategies outlined, for example, Kapelner and Bleich (2016), Chipman et al. (1998), and Pratola (2016). As outlined in the main article, we assume that marginal likelihood

$$\begin{aligned}
\Lambda(\mathcal{T}_t) &= \pi_\mathcal{T}(\mathcal{T}_t) \prod_{\ell \in \mathcal{L}_t} \left[\int \prod_{i:\boldsymbol{X}_i \rightsquigarrow (t,\ell)} \text{Normal}\{Z_i \mid \theta_0 + h_\theta(\boldsymbol{X}_i), 1\} \text{ Normal}(\theta_{t\ell} \mid 0, \sigma_\theta^2) \, d\theta_{t\ell} \right. \\
&\qquad \left. \times \int \prod_{i:Z_i > 0, \boldsymbol{X}_i \rightsquigarrow (t,\ell)} f\{Y_i \mid \boldsymbol{h}_u(\boldsymbol{X}_i), \boldsymbol{\omega}\} \, \pi_u(u_{t\ell}) \, du_{t\ell}\right] \\
&= \pi_\mathcal{T}(\mathcal{T}_t) \prod_{\ell \in \mathcal{L}_t} L_\theta(t,\ell) \cdot L_u(t,\ell).
\end{aligned}$$

can be computed in closed form. Given a transition kernel $q(\mathcal{T}_t \to \mathcal{T}_t')$ for updating $\mathcal{T}_t$, the Metropolis-Hastings acceptance probability is given by

$$A(\mathcal{T}_t \to \mathcal{T}_t') = \frac{\Lambda(\mathcal{T}_t') q(\mathcal{T}_t' \to \mathcal{T}_t)}{\Lambda(\mathcal{T}_t) q(\mathcal{T}_t \to \mathcal{T}_t')}.$$

Our choice of $q(\cdot)$ is a mixture of three possible moves: a `Birth` proposal, a `Death` proposal, and a `Change` proposal. The `Birth` step consists of the following steps.

1. Select a leaf node $\ell$ to become a branch.

2. Select a predictor $j$ to construct the split, according to $s$.

3. Sample $C_b \sim \text{Uniform}(L_j, U_j)$, with $(L_j, U_j)$ defined as in Section 2.1.

By a retrospective-sampling argument, a valid transition probability associated to this move is given by

$$q(\mathcal{T}_t \to \mathcal{T}_t') = \frac{p_{\texttt{Birth}}(\mathcal{T}_t)}{L_t}$$

where $p_{\texttt{Birth}}(\mathcal{T}_t)$ is the user-specified probability of proposing a `Birth` move (which may depend on the tree structure, as `Death` moves are not possible from the root).

The inverse of the `Birth` transition is a `Death` transition. This requires selecting the node $\ell$, which is a branch node in $\mathcal{T}_t'$. This occurs with probability

$$q(\mathcal{T}_t' \to \mathcal{T}_t) = \frac{p_{\texttt{Death}}(\mathcal{T}_t')}{B+1}$$

where $B$ is the number of branches which are not grandparents (i.e., both children are leaves).

The `Death` transition involves the following steps.

1. Select a branch node $b$, which is not a grandparent.

2. Delete the two child nodes of $b$, making it a leaf.

We again have the following forward/backward transition probabilities

$$q(\mathcal{T}_t \to \mathcal{T}_t') = \frac{p_{\texttt{Death}}(\mathcal{T}_t)}{B} \qquad \text{and} \qquad q(\mathcal{T}_t' \to \mathcal{T}_t) = \frac{p_{\texttt{Death}}(\mathcal{T}_t)}{L_t - 1}.$$

Finally, the `Change` transition is carried out as follows:

1. Select a branch node $b$ which is not a grandparent.

2. Select a new predictor $j$ according to $s$.

3. Sample a new cut point $C_b \sim \text{Uniform}(L_j, U_j)$.

As noted by Kapelner and Bleich (2016), the transition probability simplifies substantially in this case as

$$A(\mathcal{T}_t \to \mathcal{T}_t') = \frac{\Lambda(\mathcal{T}_t')}{\Lambda(\mathcal{T}_t)} \wedge 1.$$

## S.4 Proof of Lemma 1

*Proof.* To show the mapping is surjective, for any $F \in \mathscr{M}$, we can take $\pi(\boldsymbol{x}) = F_{\boldsymbol{x}}(\{0\})$ and $G_{\boldsymbol{x}}(A) = F_{\boldsymbol{x}}(A \cap \{0\}^c)/[1 - \pi(\boldsymbol{x})]$ when $F_{\boldsymbol{x}}$ has an atom at 0, and take $\pi(\boldsymbol{x}) = 0$ and $G_{\boldsymbol{x}} = F_{\boldsymbol{x}}$ otherwise.

To show the mapping is injective, consider any $(\pi, G) \neq (\pi', G')$ in $\mathscr{P} \times \mathscr{G}$, and define $F_{\boldsymbol{x}} = \pi(\boldsymbol{x})\delta_0 + [1 - \pi(\boldsymbol{x})]G_{\boldsymbol{x}}$. Define $F'_{\boldsymbol{x}}$ similarly. First, if $\pi \neq \pi'$ then there exists an $\boldsymbol{x}$ such that $F_{\boldsymbol{x}}(\{0\}) \neq F'_{\boldsymbol{x}}(\{0\})$. Conversely, suppose $\pi = \pi'$ but $G \neq G'$. Then there exists a set $A$ and a $\boldsymbol{x}$ such that $G_{\boldsymbol{x}}(A) \neq G'_{\boldsymbol{x}}(A)$; because $G$ and $G'$ do not have atoms at 0, we can assume without loss of generality that $0 \notin A$. But then, noting that $1 - \pi(\boldsymbol{x}) \neq 0$, $F_{\boldsymbol{x}}(A) = [1 - \pi(\boldsymbol{x})]G_{\boldsymbol{x}}(A) \neq [1 - \pi(\boldsymbol{x})]G'_{\boldsymbol{x}}(A) = F'_{\boldsymbol{x}}(A)$, so $F \neq F'$. □



\end{document}